\documentclass[prb,amsmath,twocolumn,showkeys,showpacs]{revtex4}
\usepackage{graphicx}
\usepackage{dcolumn}
\usepackage{amssymb}
\usepackage{bm}
\begin{document}
\newcommand{\be}{\begin{equation}}
\newcommand{\ee}{\end{equation}}
\newcommand{\caH}{\mathcal{H}}
\newcommand{\la}{\langle}
\newcommand{\ra}{\rangle}
\newcommand{\e}{\epsilon}
\newcommand{\half}{\frac{1}{2}}
\title{The influences of long range Coulomb interaction on the electronic Mach-Zehnder interferometer of quantum Hall edge states}
\author{Hyun C. Lee}
\affiliation{Department of Physics and Basic Science Research Institute,
 Sogang University, Seoul, Korea}
\date{\today}
\begin{abstract}
The influences of long range Coulomb interaction (LRCI) on
Mach-Zehnder interferometer (MZI) constructed on quantum Hall edge states  is studied employing  bosonization method. The interaction of interchannel zero-modes is shown to give rise to a characteristic energy scale which is of the order of the period of experimentally observed lobe pattern of visibility. 
The nonmonotonic behavior of visibility as found Chalker \textit{et al.} 
is understood analytically using asymptotic analysis.
\end{abstract}
\maketitle
\textit{Introduction}-
An electronic MZI is a type of matter wave interferometer that has been realized with  (integer) QH edge systems.\cite{ji,neder06,neder07}
Among many interesting phenomena observed in electronic MZI we focus on the puzzling 
lobe pattern of the interference visibility of differential conductance $\sigma(V)$ 
($V$ is a bias voltage).\cite{neder06,rou07}

This lobe pattern is hard to understand in the framework of non-interacting electrons so
 it is generally believed to be due to  many-body interaction.
A few theoretical proposals have been made to explain this lobe pattern:
the introduction of additional edge modes \cite{suk07,suk08}, the decoherence and finite temperature effect\cite{gefen07}, the shot-noise effects\cite{sim08}.
The visibility of  MZI of fractional QH edges which does not include LRCI has been also studied recently.\cite{averin}

In this Brief Report we generalize  the approach of Ref.[\onlinecite{gefen07}] in two ways:
(1)  LRCI of the zero-modes (see below) \textit{between} two channels which comprise MZI
is included (2)  both the integer QH states and the Laughlin fractional QH states
at filling fraction $\nu=1/(2n+1), n =1,2,\ldots$ are considered. In the fractional case we have to 
take both the fractional quasiparticle and the electron tunneling into account at point contacts.

A possible relevance of interchannel LRCI may be argued as follows.
The length scale of MZI in Ref.[\onlinecite{neder06}] is about $ R \sim 5 \mu \mathrm{m}$. 
Taking the typical
dielectric constant of QH devices to be $\epsilon \sim 10$, 
the associated interchannel Coulomb energy scale  $E_c$ is $ e^2/R \epsilon  \sim 5 \times 10^{-2} \mu\mathrm{eV}$,
which is of the same order of magnitude as the observed period of lobe pattern of
visibility (see Fig.3 and Fig.4 in Ref.[\onlinecite{neder06}]).
This energy scale is well expected to depend on the geometry of MZI as well as the applied magnetic field.
The investigation of this interchannel interaction of zero-modes (defined below) requires 
a very careful treatement of the so-called Klein factors of bosonization formula, 
which constitutes the most significant part of  this report. 

The results of Ref.[\onlinecite{gefen07}], in particular, the nonmonotonic behavior of visibility as a function of bias voltage (Fig.6 of Ref.[\onlinecite{gefen07}]),  depend crucially on the asymptotic behavior of one-electron Green's function in the presence of many-body interactions. 
We analyze the asymptotic behavior of Green's function
employing the method of asymptotic analysis \cite{bleistein}
, thus providing more \textit{analytic} understanding of the results.
It turns out that the momentum dependence of interaction matrix element plays an important role as we will discuss.
The main results of this report are 
(1) the exact time-dependent Klein factor which give rise to interchannel Coulomb energy scale $E_c$,
Eq.(\ref{klein-e});(2) the tunneling current which incorporates the energy scale $E_c$,
Eqs.(\ref{current1},\ref{current2});(3) the analytic  form of the visibility Eq.(\ref{asym_result}).

\textit{Model of MZI of QH edges }-
The Hamiltonian which applies to both the integer  and fractional Laughlin 
QH edge state and acts  within each channel is \cite{wen,comment1}
\begin{align}
\label{h0}
\widehat{\mathsf{H}}_{\mathrm{intra}} &= \sum_{i=1,2}\, \frac{\pi v_i}{\nu} \, \int dx \, \big [\rho_i(x)  \big ]^2 \cr
&+\frac{1}{2} \int dx dy \, V(x-y)\, \sum_{i=1,2}\, \rho_i(x) \rho_i(y),
\end{align}
where $i=1,2$ is the channel index of MZI (see Fig.1 of Ref.[\onlinecite{gefen07}] for a schematic 
view of MZI).
$\rho_i(x)=\frac{\widehat{N}_i}{L_s} + \frac{1}{2\pi}\,\partial_x \phi_i$ is the density  operator of edge $i$.
$\widehat{N}_i$ is the number operator of edge $i$ whose momentum is zero, hence it 
is often referred to as a \textit{zero-mode operator} in the context of 
bosonization.\cite{delft} $\nu$ is the filling fraction of QH system and $L_s$ is the system size.  $\phi_i$ is the boson operator which describes the collective harmonic modes with \textit{nonzero} momentum.
$v_i$ is the velocity of collective modes of channel $i$. Two channels will be assumed to be identical, so that
$v_1 =v_2 \equiv v_0$. 
$V(x) = \frac{e^2}{\epsilon \sqrt{x^2+a^2}}$ is the LRCI acting within 
each channel. $a$ is a short distance cutoff.
The fundamental commutation relation of density operators is \cite{wen}
$[ \rho_i(x), \rho_j(y) ] = i \hbar \delta_{ij} \frac{\nu}{2\pi} \partial_x \delta (x-y)$.

There are two interactions which couple two channels $i=1,2$.
One is the tunneling interaction at two quantum point contacts ($a$ and $b$) which play the role of beam splitters
of an optical interferometer.
\be
\label{tunneling}
\widehat{\mathsf{H}}_{\mathrm{t}} = t_a \Psi^\dag_1 (0) \Psi_2(0) + t_b \Psi^\dag_1(x=l_1) \Psi_2(x=l_2) +\textrm{h.c},
\ee
where the operator $\Psi_i(x)$ can be either electron or quasiparticle operator depending on the
character of the point contacts. The bosonized expression of these  operators are given in
Eqs.(\ref{boso-e},\ref{boso-q}). The tunneling amplitudes depend on the enclosed flux $\Phi$
via $ t_a t_b^* = \vert t_a t_b^* \vert e^{i \Phi/\Phi_0}$, where $\Phi_0$ 
is the flux quantum.\cite{gefen07}

The other interaction which can couple two channels is LRCI between two  channels.
For simplicity  we consider the interaction between zero-modes only. This is because 
the interchannel LRCI becomes singular logarithmically in the zero momentum limit (the singularity is cut by
the finite system size). The Hamiltonian for the
interchannel LRCI is taken to be
\be
\label{inter}
\widehat{\mathsf{H}}_C = E_C \, \widehat{N}_1 \widehat{N}_2.
\ee
Extracting the zero-mode parts from Eq.(\ref{h0}) and Eq.(\ref{inter}) we can define
a Hamiltonian for zero-modes only.
\be
\label{zero}
 \widehat{\mathsf{H}}_{\mathrm{zero}} = \frac{\pi v_0}{L_s \nu} \sum_i \widehat{N}_i^2+
E_{\mathrm{intra}} \sum_i \widehat{N}_i^2 + E_c \widehat{N}_1 \widehat{N}_2.
\ee
$E_{\mathrm{intra}}$ is an energy scale from the intra channel Coulomb intraction (the second 
term of Eq.(\ref{h0})).
The chemical potential of each channel is determined by the average number of electrons in each channel
$\la \widehat{N}_i \ra$.\cite{gefen07}
Let us define an operators $\delta \widehat{N}_i$ which describes the fluctuation of 
electron number around the average value.
\be
\label{fluctuation}
\delta \widehat{N}_i \equiv \widehat{N}_i - \la  \widehat{N}_i \ra, \;\;
\la \delta \widehat{N}_i  \ra =0.
\ee
The Hamiltonian for the fluctuations $\delta \widehat{N}_i$  can be obtained by inserting Eq.(\ref{fluctuation}) into Eq.(\ref{zero}) and discarding constant terms.
\be
\label{expanded}
\widehat{\mathsf{H}}_0 =  \sum_{i=1,2} \mu_i \delta \widehat{N}_i + E_C \delta \widehat{N}_1 \delta \widehat{N}_2,
\ee
where a contribution from $E_{\mathrm{intra}} \sum_i \widehat{N}_i^2$ is neglected since the
relative fluctuations between two channels will  play a more important role in interference.
As will be shown below, with the form of Eq.(\ref{expanded}), the time evolution of 
Klein factors can be determined \textit{exactly}.

\textit{Bosonization of electron and quasiparticle operators}- The bosonized expression for the \textit{electron} operator in the edge state of Laughlin quantum Hall liquid at filling 
fraction $\nu$ is given by\cite{wen,delft,comment2} ($i=1,2$)
\be
\label{boso-e}
\Psi_{{\rm e},i}(x) = \frac{1}{(2 \pi a)^{1/2}}\,F_{{\rm e},i} \,e^{-i 2 \pi \widehat{N}_i x /L_s \nu}\,
e^{-i \phi_i(x)/\nu},
\ee
where $F_{{\rm e},i}$ is the Klein factor which implements the fermi statistics of electron
operators of \textit{different} species. It satisfies the following relations.
\begin{align}
\label{property1}
& [\widehat{N}_i, F_{{\rm e},j}]=-F_{{\rm e},j} \delta_{ij},\quad F_i F_i^\dag = 1, \cr
&\{ F_{\mathrm{e},i}, F_{\mathrm{e},j} \} =0, \quad \{ F_{\mathrm{e},i}, F_{\mathrm{e},j}^\dag \} =0
, \; i \neq j.
\end{align}
Using the bosonization formula of  Ref.[\onlinecite{polchinski}] (with  appropriate changes of notations)
 Eq.(\ref{property1}) can be explictly realized as follows:
\begin{align}
\label{realization1}
 F_{{\rm e},1}&= e^{ i \pi (\widehat{N}_1 + \widehat{N}_2)/2 \nu}\,e^{-i \theta_1/\nu}, \cr
F_{{\rm e},2}&= e^{ -i \pi (\widehat{N}_1 + \widehat{N}_2)/2 \nu}\,e^{-i \theta_2/\nu},
\end{align}
where $\theta_{1,2}$ are the operators with zero momentum which are dual to $\widehat{N}_i$ in the following 
sense.
\be
\label{theta}
[\theta_i, \widehat{N}_j]= +i \nu \delta_{ij}.
\ee

The bosonized expression of the quasiparticle operator is 
\be
\label{boso-q}
\Psi_{{\rm q},i}(x) = \frac{1}{(2 \pi a)^{1 /2}}\,F_{\mathrm{q},i} \,e^{-i 2 \pi \widehat{N}_i x /L_s }\,
e^{-i \phi_i(x)}.
\ee
We could not find the explicit expression for the Klein factors of quasiparticle operator in literature.
Based on the idea that the quasiparticle behaves like a fraction of an electron,
we can take the $(1/\nu)$-th root (recall $1/\nu$ is an integer) of  Eq.(\ref{realization1}), thus leading to:
\begin{align}
\label{realization2}
 F_{\mathrm{q},1} &= e^{ i \pi (\widehat{N}_1 + \widehat{N}_2)/2 }\,e^{-i \theta_1}, \cr
F_{\mathrm{q},2} &= e^{ -i \pi (\widehat{N}_1 + \widehat{N}_2)/2}\,e^{-i \theta_2}.
\end{align}
The validity of Eq.(\ref{realization2}) can be confirmed by the fact that
 $F_{\mathrm{q},1}$ and $F_{\mathrm{q},2}$ satisfy the following commutation relations of \textit{fractional statistics}.
\be
\label{property2}
 [\widehat{N}_i, F_{{\rm q}, j} ] = - \nu\, F_{{\rm q}, j}\,\delta_{ij},\;\;
 F_{\mathrm{q},1} F_{\mathrm{q},2}= e^{-i \pi \nu}  F_{\mathrm{q},2} F_{\mathrm{q},1}.
\ee

\textit{Time evolution of Klein factors}-
The time evolution of Klein factors under the action of the Hamiltonian Eq.(\ref{expanded}) can
be determined exactly. We use the following operator identity.\cite{delft}
Let $A,B,D$ be some operators satisfying 
$[A,B] = D B$, $[A,D]=[B,D]=0$. Then for arbitrary function  $f(A)$ of the operator $A$, we have
\be
\label{identity}
f(A) B = B f(A+D).
\ee
Idenftifying
\be
A \to \widehat{\mathsf{H}}_0,\;\; B \to F_{e,1},\;\; D = -\mu_1 - E_c \delta \widehat{N}_2
\ee
we find (with $f(A) = e^{i A t}$)
\be
\label{klein-e}
F_{\mathrm{e},1}(t) = F_{\mathrm{e},1}(t=0) \, e^{-i \mu_1 t - i E_c \delta \hat{N}_2 t}.
\ee
The dependence of $F_{\mathrm{e},1}(t)$ on $\delta \hat{N}_2$
gives rise to \textit{additional} time dependence for correlation functions, 
which is not present in Ref.[\onlinecite{gefen07}]. This time dependence will make the visibility
exhibit features around the energy scale $E_c$ [see Eq.(\ref{visibility})].
Similarly,
\be
F_{e,2}(t) = F_{e,1}(t=0) \, e^{-i \mu_2 t - i E_c \delta \widehat{N}_1 t}.
\ee
As for quasiparticle operators, we have 
\begin{align}
\label{klein-q}
 F_{\mathrm{q},1}(t) &= F_{\mathrm{q},1}(t=0) \, 
e^{-i \nu \mu_1 t - i \nu E_c \delta \widehat{N}_2 t}, \cr
F_{\mathrm{q},2}(t) &= F_{\mathrm{q},2}(t=0) \, e^{-i \nu \mu_2 t - i \nu E_c \delta \widehat{N}_1 t}. 
\end{align}

\textit{Tunneling Currents}-
The tunneling current operator is given by (electron tunneling is assumed)
\begin{align}
&\widehat{I} = (-e)\, \frac{d \widehat{N}_2}{d t} = \left( -\frac{ie}{\hbar} \right)\, 
\big[ \widehat{H}_{\mathrm{total}},
\widehat{N}_2 \big ] \cr
&=\left( -\frac{ie}{\hbar} \right)\,\Big( t_a \Psi_1^\dag(0) \Psi_2(0) + t_b \Psi_1^\dag(l_1) \Psi_2(l_2)
-\mathrm{h.c.} \Big ),
\end{align}
where $\widehat{H}_{\mathrm{total}}$ is the sum of Eqs.(\ref{h0},\ref{tunneling},\ref{inter}).
A straightforward time-dependent perturbation in $t_{a,b}$ yields\cite{gefen07}
\begin{align}
\label{current}
I& \equiv  \la \hat{I}(t=0) \ra =-\frac{e}{\hbar}\, \int_{-\infty}^0 dt
 \Big( \sum_{k=1}^4 X_k (t) + \mathrm{h.c.} \Big ), \cr
X_1(t) &= \vert t_a \vert^2 \la [ \Psi_1^\dag(0,0) \Psi_2(0,0), \Psi_2^\dag(0,t) \Psi_1(0,t) ] \ra, \cr
X_2(t) &= \vert t_b \vert^2 \la [ \Psi_1^\dag(l_1,0) \Psi_2(l_2,0), \Psi_2^\dag(l_2,t) \Psi_1(l_1,t) ] \ra, \cr
X_3(t)&=t_a t_b^* \la [ \Psi_1^\dag(0,0) \Psi_2(0,0), \Psi_2^\dag(l_2,t) \Psi_1(l_1,t) ] \ra, \cr
X_4(t)&=t_a^* t_b \la [ \Psi_1^\dag(l_1,0) \Psi_2(l_2,0), \Psi_2^\dag(0,t) \Psi_1(0,t) ] \ra, 
\end{align}
where the expectation value should be taken with respect to the eigenstates of the Hamiltonian
$\mathsf{H}_{\mathrm{intra}}+\mathsf{H}_{C}$ [Eqs.(\ref{h0}, \ref{inter}].
For the  quasiparticles, the current $I$ must be multiplied by $\nu$.
The visibility of interference originates from the correlation functions $X_3(t)$ and $X_4(t)$.

Each correlation function $X_i(t)$ can be factorized into 
the product of the nonzero-mode (also called oscillators) contribution and the zero-mode contribution.
Furthermore  the nonzero-mode contribution is a product of that of channel 1 and channel 2, since
the Hamiltonian of the nonzero-modes of channel 1 and 2 commutes each other. 
The nonzero-mode contribution is identical with that of Ref.[\onlinecite{gefen07}], and the
details are omitted.
The zero-mode contribution, however, cannot be expressed as a product of contribution from each channel,
since the zero-modes of two channel are coupled by the Hamiltonian Eq.(\ref{inter}).

In spite of this noncommutativity, 
the zero-mode contribution can be computed exactly owing to the results Eqs.(\ref{klein-e},\ref{klein-q}).
For example, the zero-mode contribution of $X_1(t)$ in the case of 
the quasiparticle tunneling is given by
\be
\label{zero-mode1}
\la F^\dag_{\mathrm{q} 1}(0) F_{\mathrm{q} 2}(0) F^\dag_{\mathrm{q} 2}(t) F_{\mathrm{q} 1}(t) \ra
=e^{i \nu ( \mu_2 - \mu_1) t} e^{-i \nu^2 E_c t}.
\ee
To obtain Eq.(\ref{zero-mode1}) we insert the results Eq.(\ref{klein-e},\ref{klein-q}) for
the time dependent Klein factors, then employ the commutation relation between Klein factor and the
zero-mode operator $\delta \widehat{N}_i$, and use $\la \delta \widehat{N}_i \ra =0$.

Using complex conjugation properties of  $X_i$ it can be shown that
($I_0$ and $I_\Phi$ defined in an obvious way)
\be
\label{current-integral}
I =I_0+I_\Phi=-\frac{e}{\hbar}\, \int_{-\infty}^\infty dt \Big[ Y_0(t)+2 \mathrm{Re} Y_\Phi(t) \Big ].
\ee
The visibility is basically determined by the flux depedendent conductance 
$ \sigma_\Phi = d I_\Phi / d V$ ($eV= \mu_2-\mu_1$).
For  electron tunneling, the flux independent part is 
\begin{align}
\label{current1}
Y_{0}^{(e)}(t) &= \frac{1}{ (2\pi a)^2}\, (\vert t_a \vert^2+ \vert t_b \vert^2) e^{i  (\mu_2 -\mu_1) t} \cr
&\times \Big[ g_{e}^2(x=0,-t) e^{-i  E_c t} - \mathrm{h.c.} \Big ],
\end{align}
and the flux dependent part is 
\begin{align}
\label{current2}
&Y_{\Phi}^{(e)}(t) = \frac{t_a t_b^*}{(2\pi a)^2}\,e^{i  (\mu_2 - \mu_1) t} \,
e^{ 2\pi i ( \la N_2 \ra l_2 - \la N_1 \ra l_1)/ \nu L_s} \cr
&\times \Big[ g_e(-l_1,-t) g_e(-l_2,-t) e^{-i E_c t} 
- \mathrm{h.c.}  \Big ].
\end{align}
$g_e(x,t)$ is the nonzero-mode contribution for the one-electron Green's function, which has been obtained in
Ref.[\onlinecite{gefen07}]. At $T=0$,
\begin{align}
\label{cor:e}
 g_{e}(x,t) &= \exp\Big[-c_e(x,t) - i s_e(x,t) \Big ], \cr
c_e(x,t) &= \frac{1}{\nu} \int_0^\infty \frac{dq}{q}\, e^{-a q} 
\Big( 1- \cos( q x + \omega_q t) \Big ), \cr
s_e(x,t) & = \frac{1}{\nu} \int_0^\infty \frac{dq}{q}\,e^{-a q} \sin (q x + \omega_q t),
\end{align}
where the frequency of nonzero-modes is given by
\be
\label{frequency}
\omega_q = q [ v_0 + v_c \, \ln \frac{\zeta}{q a} ],\;\; v_c = \frac{\nu e^2}{2\pi \hbar \epsilon},
\ee
where $qa \lesssim 1$ is assumed and  $\zeta \approx 1.13$ is a numerical constant.
The logarithmic factor of Eq.(\ref{frequency}) comes from  
the intrachannel Coulomb interaction matrix element.
The major difference from those of Ref.[\onlinecite{gefen07}] is the presence of 
the factor $e^{\pm i E_c t'}$  which originates
from the interchannel LRCI.
For quasiparticle tunneling, the following modifications are to be made
\begin{align}
& E_c \to \nu^2 E_c, \;\; (\mu_2 -\mu_1) \to \nu (\mu_2-\mu_1), \cr
& c_e(x,t) \to \nu^2 c_e(x,t), \;\; s_e(x,t) \to \nu^2 s_e(x,t).
\end{align}

We note that the energy scale $E_c$ enters in such a way it is not a mere additive renormalization of chemical potential.
Also, in higher $2n$-th order expansion in tunneling amplitude $t_{a},t_{b}$, 
a factor $e^{\pm i n E_c t'}$ will emerge since this factor is generated by commutation of Klein factors.
Recall that the Klein factors accompany the tunneling amplitude in Eq.(\ref{tunneling}). 
Thus, one can expect  some features   of visibility which is due to $g_e(x,t)$ will appear 
being centered at energies $ n E_c$ with decreasing magnitude.
In the second order expansion of $t_{a,b}$, the visibility (or equivalently flux dependent
conductance $\sigma_\Phi$) is found to exhibit nonmonotonic behavior.\cite{gefen07}
Let us try to  understand the origin of such behavior in a more analytic way.

\textit{Asymptotic behavior of one-electron Green's function}-
The correlation function $g_{e}(x,t)$ cannot be evaluated in a closed form, so that
an analytic form of asymptotic behavior will be of great help in understanding the nonmonotonic
dependence of conductance on bias.
First of all we note that the equal time Green's function $g_{e}(x,t=0)$
is independent of interactions. 
However, we are mostly interested in the long time limit, 
so that we  focus on the domain $ \vert x / v_c t \vert \ll 1 $.
We write  $g_{e}(x,t)$ in the following form:
\begin{align}
\label{asymp1}
g_{e}(x,t) &= \exp\Big[- \mathrm{const.} + I(x,t) \Big ], \cr
I(x,t) &= \frac{1}{\nu}\int_0^\infty \frac{dq e^{-aq}}{q} e^{-i \varphi_q(x,t)},
\end{align}
where $\mathrm{const.}=\frac{1}{\nu}\int_0^\infty \frac{dq e^{-aq}}{q}$ is a (infinite) constant, and
\be
\varphi_q(x,t) = q x + \omega_q t
\ee
is a phase function.
In fact, the integral $I(x,t)$ diverges at $q=0$, and the divergence   is cancelled by the above constant.
In spite of this divergence,  the form of Eq.(\ref{asymp1}) is more preferable 
for the asymptotic analysis.
To avoid the divergence at $q=0$, we employ the technique of dimensional regularization\cite{ramond}:
replace the factor $1/q$ of Eq.(\ref{asymp1}) by $ 1/q^{1-\alpha}$ with $\alpha > 0$, and take
the limit $\alpha \to 0$ limit and extract the finite contribution.
An infinity which appears in extraction is cancelled by the infinite constant mentioned above.

In the long time /distance limit, the phase $\varphi(x,t)$ becomes very large in
the generic domain, then  we can use the stationary phase approximation.\cite{bleistein}
In general, there exist two contributions to the asymptotic behavior of the integral of the type 
of $I(x,t)$: 
one from the ends of integral interval ($q=0$ and $q=\infty$, clearly the contribution from
$q=\infty$ is negligible) and the other
from the stationary point(s) where $d \varphi_q(x,t)/d q =0$.

The contribution from the end point at $q=0$ can be obtained by integration by parts method. 
\cite{bleistein}
Define a variable $u$ as follows: 
\be
u=u(q) =  q \frac{x}{t} + \omega_q =q v_c \ln (\frac{\eta}{qa}),
\ee
where $\eta= \zeta e^{[(x/t) + v_0]/v_c}$.
Changing the integration variable from $q$ to $u$, 
 the relevant integral becomes
\be
\int_0^\infty d u \vert\frac{d q}{d u} \vert \frac{e^{-a q}}{[q(u)]^{1-\alpha}} e^{ - i t u}.
\ee
Let us consider the case where $u > 0$.
Within logarithmic accuracy we have $
q(u) \sim \frac{u}{v_c} [\ln \frac{\eta v_c}{u a}]^{-1}$ 
and $\frac{d q}{d u} \sim \frac{1}{v_c} \frac{1}{\ln  \eta^* v_c /u a}$ with 
$\eta^* = e^{ [(x/t)+ v_0]/v_c -1}$.
Performing the partial
integration along \textit{imaginary axis}\cite{bleistein}, taking 
$\alpha \to 0$ limit, and extracting the finite part,  we obtain
\be
\label{endpoint}
I_{\mathrm{end}}(x,t) \sim - \ln[ \frac{v_c t}{a}\, \ln (  \eta v_c t/a) ] -  i \, \mathrm{sign}(t)\, \pi/2 
\ee
The effect of LRCI within each channel is reflected in the double logarithmic correction.
$ \ln[ \ln ( \eta v_c t/a)]$. 
Note that this correction also depends on position $l_1,l_2$ through $\eta$.
In terms of the Green function $g_e(x,t)$,  the end point contribution 
is  roughly $1/[ t \ln t]^{1/\nu}$,  which evidently  cannot cause the nonmonotonic behavior of 
flux dependent conductance $\sigma_\Phi$ as shown in Ref.[\onlinecite{gefen07}].

Next we turn to the contribution from stationary point.
The condition for the stationary  point is
\be
\label{condition}
\frac{d \varphi_q(x,t)}{d q}\Big \vert_{q = q_c} =0 \to
-\frac{x}{t} =  \frac{d \omega_q}{d q} \vert_{q =q_c} \sim v_c \ln \frac{\bar{\zeta}}{q_c a},
\ee
where $\bar{\zeta} = \zeta e^{(v_0/v_c) -1}$, and $q_c$ is determined by $x/t$.
For the condition Eq.(\ref{condition})
 to be satisfied for very long time limit (namely $\vert x/v_c t \vert \ll 1$)
there should be a point where $   \frac{d \omega_q}{d q} \vert_{q =q_c} =0$. In other words,
the frequency should attain a local maximum or mininum at finite momentum.
Therefore if $\omega_q$ is a monotonic function of $q$, then  there will be 
no stationary point, so that the nonmonotonic behavior of visibility would not appear.
 Another dispersion $ \omega_q = v_0 q - b q^3$  with $ b>0$ \cite{gefen07} which also 
shows nonmonotous behavior satisfies the local maximum condition.
 It is easily seen $
\frac{d^2 \varphi_q(x,t)}{d q^2} \vert_{q=q_c} = -\frac{v_c}{q_c}$
Then the standard stationary phase approximation \cite{bleistein} gives
\be
\label{stationary}
I_{\mathrm{sta}}(x,t) \sim \sqrt{ \frac{ 2 \pi}{\vert t \vert   q_c v_c}} e^{-i \frac{\pi}{4} \mathrm{sgn} t} e^{-i  \varphi_q(x,t) \vert_{q=q_c}}.
\ee
In the long time limit $\vert I_{\mathrm{end}}(x,t) \vert \gg \vert I_{\mathrm{sta}}(x,t) \vert$.
Were it not for the $1/q$ singularity at $q=0$, the stationary point contribution ($1/\sqrt{t}$) would dominate the singularity-free end point contribution ($1/t$). 
Eq.(\ref{stationary}) explains the oscillating behavior of Green function at long time tail as 
found in Fig. 4 of  Ref.[\onlinecite{gefen07}].
Now  the asymptotic form of the correlation function $g_e(x,t)$ is given by
\be
\label{asym_result}
g_e  \sim e^{I_{\mathrm{end}}+I_{\mathrm{sta}}} \sim \frac{1}{[ \frac{v_c t}{a}\, \ln ( \bar{ \eta} v_c t/a)]^{1/\nu}}\,
\big(1+ I_{\mathrm{sta}} \big ).
\ee


\textit{Discussions}-
The origin of the nonmonotonic behavior of visibility found in Ref.[\onlinecite{gefen07}] is the 
oscillating tail of one-particle eletron Green function. In this report we have found that the oscillating 
tail is due to the specific momentum dependence of collective excitation.
The investigation of  the detailed form of the visibility requires the numerical integration\cite{gefen07},
but the essential features can be understood qualitatively using Eq.(\ref{asym_result}).
Taking the typical time as $ t \sim 1/\vert eV - E_c \vert $, ($ \Delta = \hbar v_c /a$) one can estimate
the integral of Eq.(\ref{current-integral}) to obtain
\begin{align}
\label{visibility}
\sigma_\Phi \sim  \left[  \vert e V  -E_c \vert \ln \frac{\vert e V-E_c\vert}{\Delta} \right]^{(2/\nu)-2}  \cr
\times  \Big[1+ \vert\frac{ e V - E_c}{\Delta} \vert^{1/2}\cos( \tilde{\varphi}_c(V)) \Big],
\end{align}
where $\tilde{\varphi}_c(V)$ is basically the phase function $\varphi_q$ evaluated at the stationary point, 
and it slowly varies as a function of bias. The factor $\cos( \tilde{\varphi}_c(V))$ is responsible
for the nonmonotonic behavior.
At integer filling $\nu=1$ the presence of energy scale $E_c$ is not so pronounced because the exponent
of prefactor of Eq.(\ref{visibility}) vanishes. Compared to  Ref.[\onlinecite{gefen07}], the most dramatic
difference with QH case ($\nu < 1 $, for electron tunneling) would be the strong suppression of amplitude
near $eV \sim E_c$.
As for the quasiparticle tunneling (at $\nu < 1$), the time integral of Eq.(\ref{current-integral}) diverges
in the long time limit,
which implies that MZI is entering strong coupling regime at low temperature.
It is well-known that the strong coupling regime of QH point contact is well-described by
\textit{electron tunneling} picture.\cite{imura97}
At high temperature, the divergence in long time limt is cut by $1/T$. Thus as temperature decreases,
we can expect a crossover from quasiparticle tunneling to electron tunneling, and at the same time
the interchannel Coulomb energy scale changes from $\nu^2 E_c$ to $E_c$. All these features 
can be readily checked experimentally.

We have to note that the results of both Ref.[\onlinecite{gefen07}] and the present report have been obtained in the 
second order perturbation with respect to tunneling amplitudes. As such both results do not seem to explain the 
experimental results of the \textit{periodic} behavior of visibility. However, the present report indicates the
existence of a series of energy scales $n E_c$ ($n=1,2,3,\ldots$), and it clearly suggests that we need the higher order perturbations to reveal a possible underlying periodic structure of visibility.

\begin{acknowledgements}
The author is thankful to  Hyun-Woo Lee and H.-S Sim for introducing this problem.
This work was supported by the Korea Science and Engineering Foundation (KOSEF) grant funded by
the Korea government (MEST) (No. R01-2008-000-10805-0).
\end{acknowledgements}



\begin{thebibliography}{}
\bibitem{comment1} Spins are assumed to be completely polarized.
\bibitem{comment2} We choose the convention of left-moving particles.
\end{thebibliography}
\end{document}